\documentclass[%
 reprint,
 amsmath,amssymb,
 aps,
]{revtex4-2}

\usepackage[colorlinks=true,linkcolor=blue,citecolor=blue,urlcolor=blue]{hyperref}
\usepackage{graphicx}
\usepackage{dcolumn}
\usepackage{bm}

\newcommand{\isotope}[2]{$^{\text{#2}}$#1}

\begin{document}

\preprint{APS/123-QED}
\newcommand{\physrev}{}

\title{Growth of Ultra-high Purity NaI(Tl) Crystal for Dark Matter Searches}

\author{Burkhant Suerfu}
\email{suerfu@princeton.edu}

\author{Masayuki Wada}
\altaffiliation[Now at ]{AstroCeNT, Nicolaus Copernicus Astronomical Center of the Polish Academy of Sciences, Poland}

\author{Winston Peloso}
\altaffiliation[Now at ]{University of Colorado Boulder}

\author{Michael Souza}
\altaffiliation[Also at ]{Department of Chemistry, Princeton University}

\author{Frank Calaprice}
\affiliation{Department of Physics, Princeton University\\
Princeton, New Jersey, 08544, U.S.A.
}

\author{Joshua Tower}

\author{Guido Ciampi}
\affiliation{Radiation Monitoring Devices, Inc.\\
Watertown, Massachusetts, 02472, U.S.A.
}




\date{\today}

\begin{abstract}
The annual modulation of scintillation event rate observed by the DAMA/LIBRA experiment has been a long-standing controversy in the quest of the direct detection of dark matter. The effort to 
definitively confirm or refute the annual modulation has turned out to be challenging due to the lack of NaI(Tl) crystals with high enough radio-purity. Most recently, we successfully grew a 6-kg ingot free from contamination during growth, from which a 3.4-kg crystal scintillator was made. The \isotope{K}{39} concentration in the final crystal is estimated to be 4.3$\pm$0.2~ppb, 
unprecedented for NaI(Tl) crystals. The \isotope{Pb}{210} activity is estimated to be 0.34$\pm$0.04~mBq/kg via $\alpha$ counting of \isotope{Po}{210}, among the lowest of currently-running NaI-based dark matter experiments except DAMA/LIBRA. More importantly, the techniques and protocols we have developed will further contribute to the growth of higher purity NaI(Tl) crystals for dark matter searches.

\end{abstract}


\maketitle


\section{Introduction}

The nature of dark matter is one of the most important questions of modern physics. Although most attempts at directly detecting dark matter have yielded in null results, the DAMA/LIBRA experiment operating underground at the Laboratori Nazionali del Gran Sasso~(LNGS) observes an annual modulation in event rate. Such a modulation is expected from the annual change in the relative velocity between the Earth and the dark matter halo as the Earth rotates around the Sun and the Sun moves at a constant speed in the dark matter halo~\cite{annual-modulation-theory, dl-phase2}.
The observed annual modulation has an amplitude of approximately 0.01~cpd/kg/keV on top of a constant background of about 1~cpd/kg/keV in the 2-6~keV range, and a phase consistent with the dark matter halo model~\cite{dl-phase2}.

DAMA/LIBRA utilized $\sim$250~kg of high-radiopurity NaI(Tl) crystals arranged in a 5$\times$5 array shielded by a combination of polyethylene, high-purity copper and concrete~\cite{dl-apparatus}. DAMA/LIBRA and its predecessor DAMA/NAI have been taking data for over 16 years, achieving a statistical significance of 12.9~$\sigma$~\cite{dl-phase2} for the annual modulation signal. However such an annual modulation is in tension with the results of many other direct dark matter detection experiments~\cite{Agnese2018,Agnes2018,Aprile2018}. To investigate whether the annual modulation is caused by a systematic error, unknown detector effect or non-standard dark matter interaction, an independent low-background experiment using the same target material--NaI(Tl)--with significantly lower background level is needed.

Several experiments have been trying to confirm or refute the DAMA/LIBRA annual modulation, including ANAIS, DM-Ice, KIMS and COSINE~\cite{anais,dmice,kims,cosine}. However, none of these efforts have achieved a background low enough to confirm or refute the modulation effectively within a reasonable amount of time, and the biggest source of background comes from radioactive impurities inside the crystal. In particular, \isotope{K}{40} is of a great threat since it can decay with 10.72\% branching ratio by the capture of a K-shell electron and the emission of 1.46-MeV $\gamma$-ray. When the high-energy $\gamma$-ray escapes the crystal, X-rays and Auger electrons from the subsequent 3-keV atomic transition appear in the middle of the 2-6~keV energy region of interest. In addition, 
\isotope{Pb}{210} and \isotope{H}{3} are of potential concern since these $\beta$-decay isotopes have large overlaps between their $\beta$~spectra and the region of interest.

In this article, we report our progress in developing ultra-high purity NaI(Tl) crystals, and present results of impurity measurements on NaI-033---our most recent 3.4-kg NaI(Tl) scintillating crystal.

\section{Crystal Growth}
NaI-033 was grown using ultra-high purity NaI powder.
The concentrations of \isotope{U}{238} and \isotope{Th}{232} in the powder have been pre-screened to be below 0.6~ppt and 0.5~ppt, respectively~\cite{emily-thesis}. The concentration of \isotope{K}{39} is measured to be 7~ppb by Seastar Chemicals using inductively coupled plasma mass spectrometry~(ICP-MS).  Since \isotope{K}{40} is a primordial radionuclide with a well-defined natural abundance of 0.011\%, in the rest of this article ICP-MS measurement of stable \isotope{K}{39} is used instead of direct counting of \isotope{K}{40}.


Prior to crystal growth, 6~kg of ultra-high purity NaI powder was thoroughly dried at Radiation Monitoring Devices~(RMD) to remove trace amount of water which can not only affect crystal growth adversely, but also contribute to crystal radioactivity via \isotope{H}{3} decay. The drying procedure adopted was based on precision drying techniques developed at Princeton University which used vacuum baking and purging with inert gas at several increasing steps of temperature until no more release of water can be detected~\cite{suerfu-thesis}. 

The dry NaI powder is subsequently mixed with 
high-purity thallium iodide~(TlI) powder~(99.999\%) and sealed inside a 4-in-diameter, 2-ft-long 
high-purity crucible.
To prevent contamination, the crucible is carefully cleaned with high-purity acids.


After the seal, the crucible was taken to RMD for crystal growth using the vertical Bridgman method~\cite{bridgman}. The furnace temperature profile and growth rate were adjusted to yield defect-free single crystals. The Bridgman method was chosen because the molten raw material can be completely sealed inside the crucible during crystal growth, thus eliminating the risk of contamination.

\begin{figure}[htb!]
    \centering
    \ifdefined\physrev
    \includegraphics[height=0.7\linewidth]{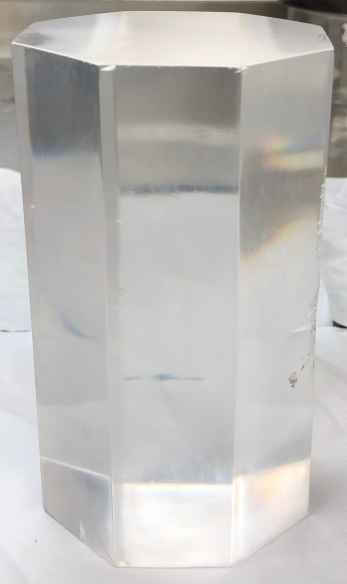}
    \else
    \includegraphics[height=0.4\linewidth]{figures/NaI033-2.png}
    \fi
    \caption{NaI-033 octagonal crystal with surface polished with high-purity isopropyl alcohol.}
    \label{fig:crystal}
\end{figure}

After the crystal growth, the ingot was cut into a 3.4~kg, 151-mm-tall octagon~(NaI-033) using a diamond wire saw. The crystal surface was subsequently chemically polished with semiconductor-grade ethanol/isopropyl alcohol to remove surface contaminants introduced during the crystal cutting~(Fig.~\ref{fig:crystal}). Such a cleaning procedure has been previously tested to have little or no influence on the crystal's light yield compared to traditional mechanical polishing~\cite{suerfu-thesis}.

\section{Detector Setup}

\begin{figure}[htb!]
    \centering
    \ifdefined\physrev
        \includegraphics[width=\linewidth]{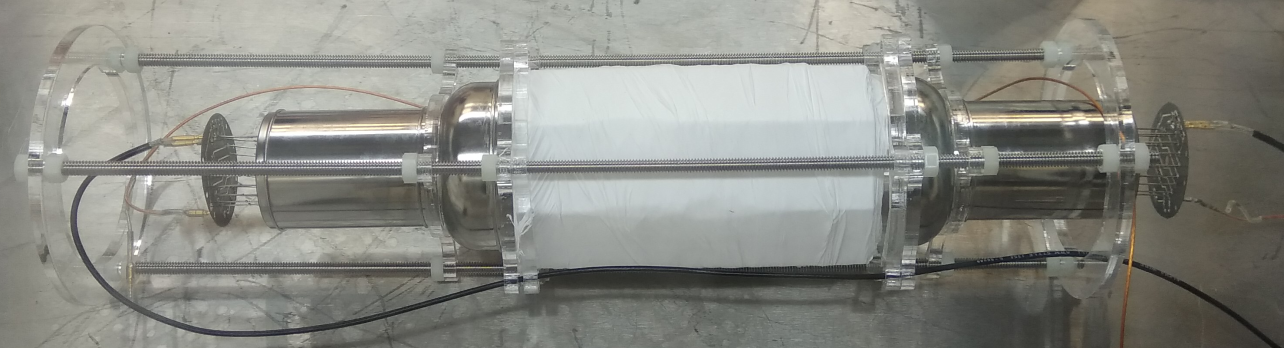}
    \else
        \includegraphics[width=0.9\linewidth]{figures/nai033_assembly.png}
    \fi
    \caption{NaI-033 crystal detector module. The crystal is wrapped with 10~layers of PTFE tape and coupled to two Hamamatsu~R11065 PMTs via optical grease. The assembly is held in place by acrylic holders, Nylon nuts and three stainless steel threaded rods.}
    \label{fig:setup}
\end{figure}

To characterize the crystal's light yield and to measure the activity of \isotope{Pb}{210} via \isotope{Po}{210}, NaI-033 was assembled into a detector module in a glovebox inside the radon-free cleanroom at Princeton~(Fig.~\ref{fig:setup}). The crystal was wrapped with approximately 10~layers of PTFE tape and optically coupled directly to Hamamatsu R11065 photomultiplier tubes on each end. The crystal-PMT assembly is held together by acrylic holders, Nylon nuts and stainless steel threaded rods. The entire structure is placed inside a 5-in-ID aluminum enclosure and operated at ground level.

During operation, the detector module is shielded from the ambient background radiation with at least 4~inches of lead in every direction. Two plastic scintillators are placed on top of the detector module as muon vetoes. To prevent the degradation of the crystal due to moisture, the detector module is constantly purged with dry boil-off nitrogen. The two PMTs are biased at negative high voltages for cleaner baselines, and the PMT bodies are insulated from metallic components with acrylic plastic. The PMT signals are digitized by a CAEN~V1720 waveform digitizer, and acquisition is triggered by cross-threshold coincidence between the two PMTs with a coincidence window of 80~ns. The DAQ software is developed using the polaris general-purpose modular DAQ framework~\cite{polaris}.

The crystal's light yield is measured by placing calibration sources next to the aluminum shell inside the lead shielding. The mean charge corresponding to single photoelectrons~(spe) is obtained by identifying and integrating trailing spe pulses after each main pulse~\cite{suerfu-thesis}. The number of photoelectrons in the photopeak is obtained by dividing the mean integral of the main pulses by that of single photoelectrons. The crystal demonstrated a light yield of 14.8$\pm$0.5~p.e./keV at the 661.7~keV from \isotope{Cs}{137}, and 16.4$\pm$0.3~p.e./keV at 59.5~keV from \isotope{Am}{241}. Such an increase of light yield towards low energy is consistent with previous literature~\cite{knoll}.

\section{Crystal Purity}

\subsection{\isotope{K}{39}}

\begin{figure}[htb!]
    \centering
    \ifdefined\physrev
        \includegraphics[width=\linewidth]{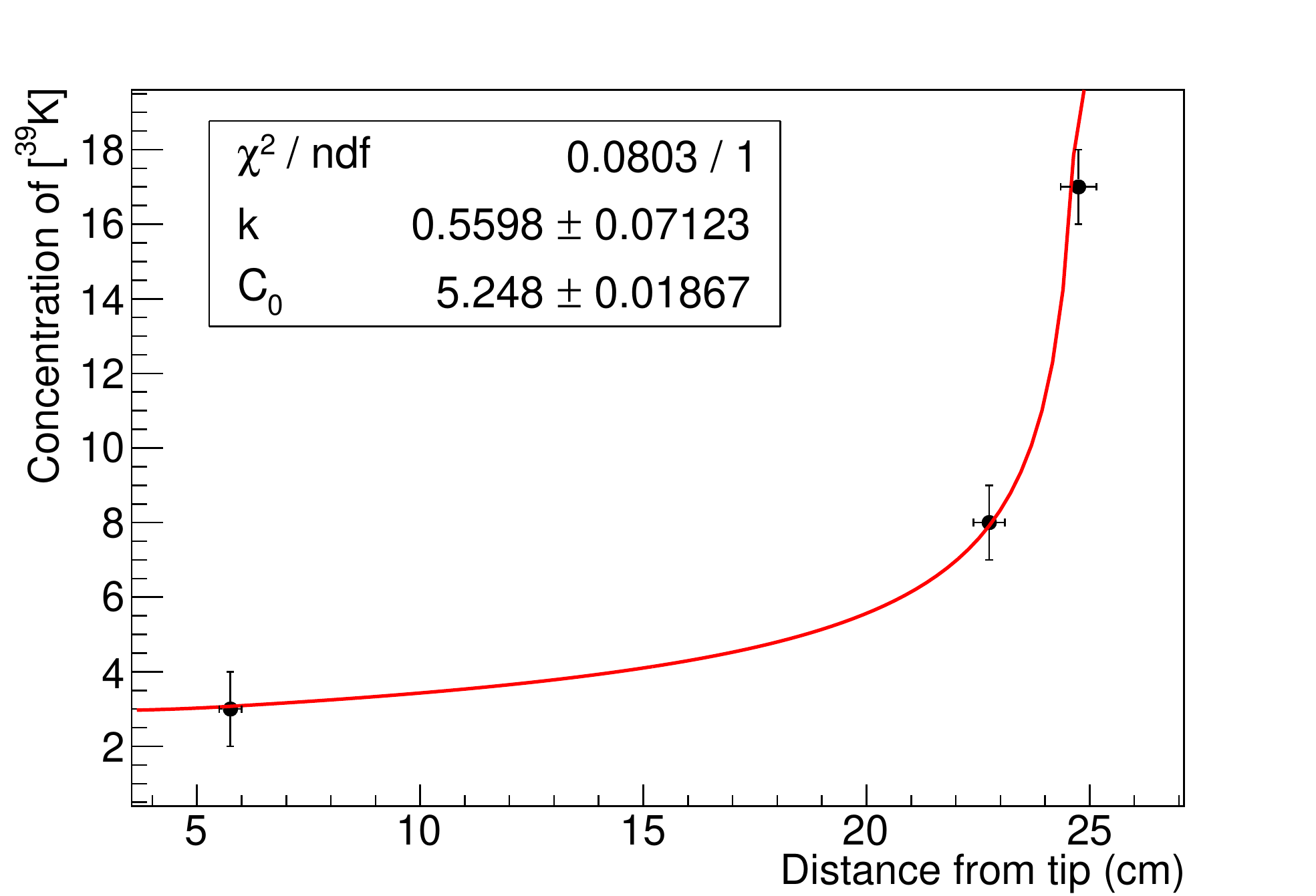}
    \else
        \includegraphics[width=0.7\linewidth]{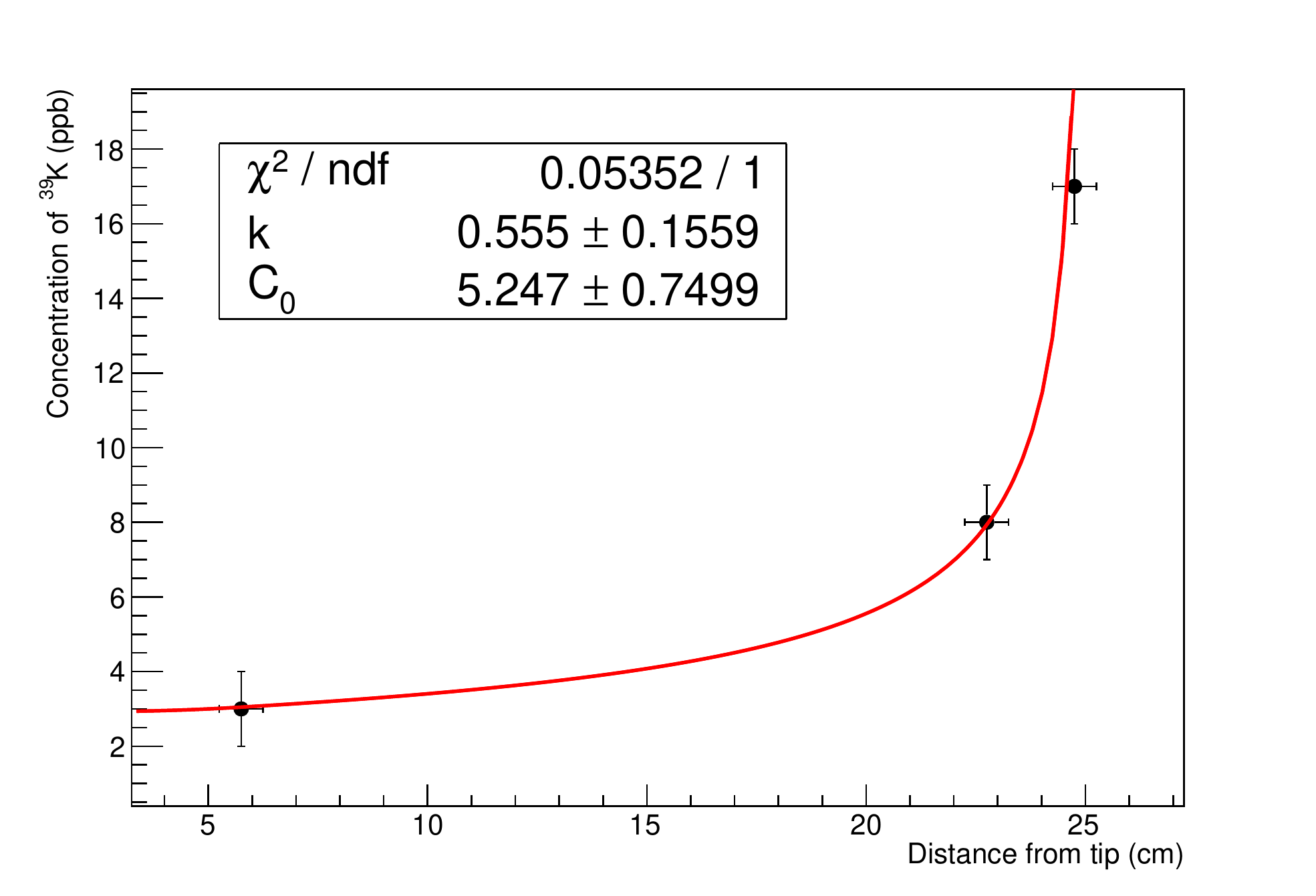}
    \fi
    \caption{Local concentration of \isotope{K}{39} as a function of distance from the tip of the ingot where crystallization begins in a vertical Bridgman process. During the crystallization, K is expelled from the crystal matrix and therefore concentrates towards the end. From the fit, the concentration of \isotope{K}{39} is estimated to be 4.3$\pm$0.2~ppb in the final crystal, obtained by averaging the fit function between 6~cm and 23~cm.}
    \label{fig:k-distrib}
\end{figure}

To determine the concentration of \isotope{K}{39} in the final crystal, three samples from the as-grown ingot were analyzed using ICP-MS at Seastar Chemicals. Prior to each measurement, the samples' surfaces were carefully cleaned with high-purity ethanol and de-ionized water multiple times to remove surface cross-contamination introduced during crystal cutting. The distribution of \isotope{K}{39} is fitted against model~\cite{suerfu-thesis}, with distribution coefficient~k and average impurity concentration~C$_0$ in the ingot as two free parameters~(Fig.~\ref{fig:k-distrib}). The distribution coefficient---defined as the ratio of impurity in the solid phase to that in the liquid phase during crystallization---is a measure of the degree of separation of the impurity from the matrix.  From the fit, the average \isotope{K}{39} concentration in the final NaI-033 crystal is estimated to be 4.3$\pm$0.2~ppb, about three times lower than 13~ppb reported in the DAMA/LIBRA crystals~\cite{dama-13ppb}.

\subsection{\isotope{Pb}{210}}

\begin{figure}[htb!]
    \centering
    \ifdefined\physrev
  	    \includegraphics[width=\linewidth]{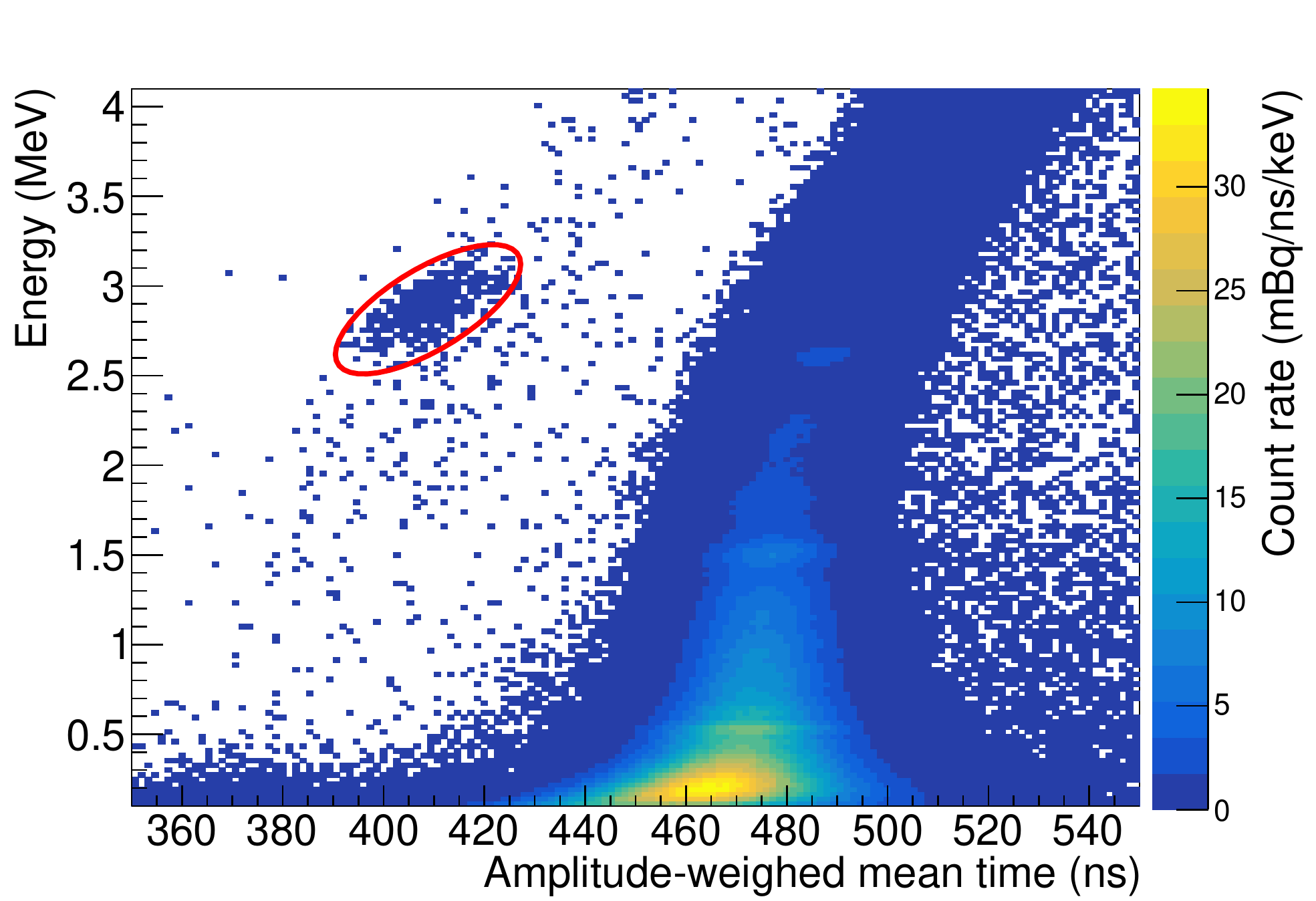}
  	\else
  	    \includegraphics[width=0.7\linewidth]{figures/nai033_alpha_psd_21.pdf}
  	\fi
    \caption{Amplitude-weighed mean time v.s. energy for NaI-033. The energy scale is calibrated with the 2.6-MeV $\gamma$-ray from \isotope{Tl}{208}. The region of interest for \isotope{Po}{210} $\alpha$ counting is indicated with a red solid circle around 2.9~MeV and 410~ns. 
    }
    \label{fig:po-discrim}
\end{figure}

Direct measurement of \isotope{Pb}{210} requires accurate measurement of the $\beta$ spectrum of \isotope{Pb}{210} or \isotope{Bi}{210}, which is difficult without being underground with sufficient shielding. However, the 5.3-MeV $\alpha$ particle from \isotope{Po}{210} decay has very short range and is relatively high in energy. By exploiting pulse-shape discrimination of NaI(Tl), $\alpha$ decays can be effectively identified and counted while on the Earth's surface~\cite{dl-apparatus}. Fig.~\ref{fig:po-discrim} shows a 2D color plot of energy v.s. pulse-shape parameter for the first 6 days. In this plot the amplitude-weighed mean time is chosen as the pulse-shape parameter. It is defined as

\begin{equation}
    \langle\tau\rangle=\frac{\sum_{i}A_i \tau_i}{\sum_{i}\tau_i},
\end{equation}

where $A_i$ is the height of the scintillation pulse at time $\tau_i$ and summation is over 1.2~$\mu$s since the start of the pulse.

To determine the activity of \isotope{Pb}{210}, background data of approximately 80 days were taken 3 months after the crystal growth to account for the ingrowth of \isotope{Po}{210}, and data within each 6-day period are grouped together to obtain the time-dependence of \isotope{Po}{210} activity. 

\begin{figure}[htb!]
    \centering
    \ifdefined\physrev
        \includegraphics[width=\linewidth]{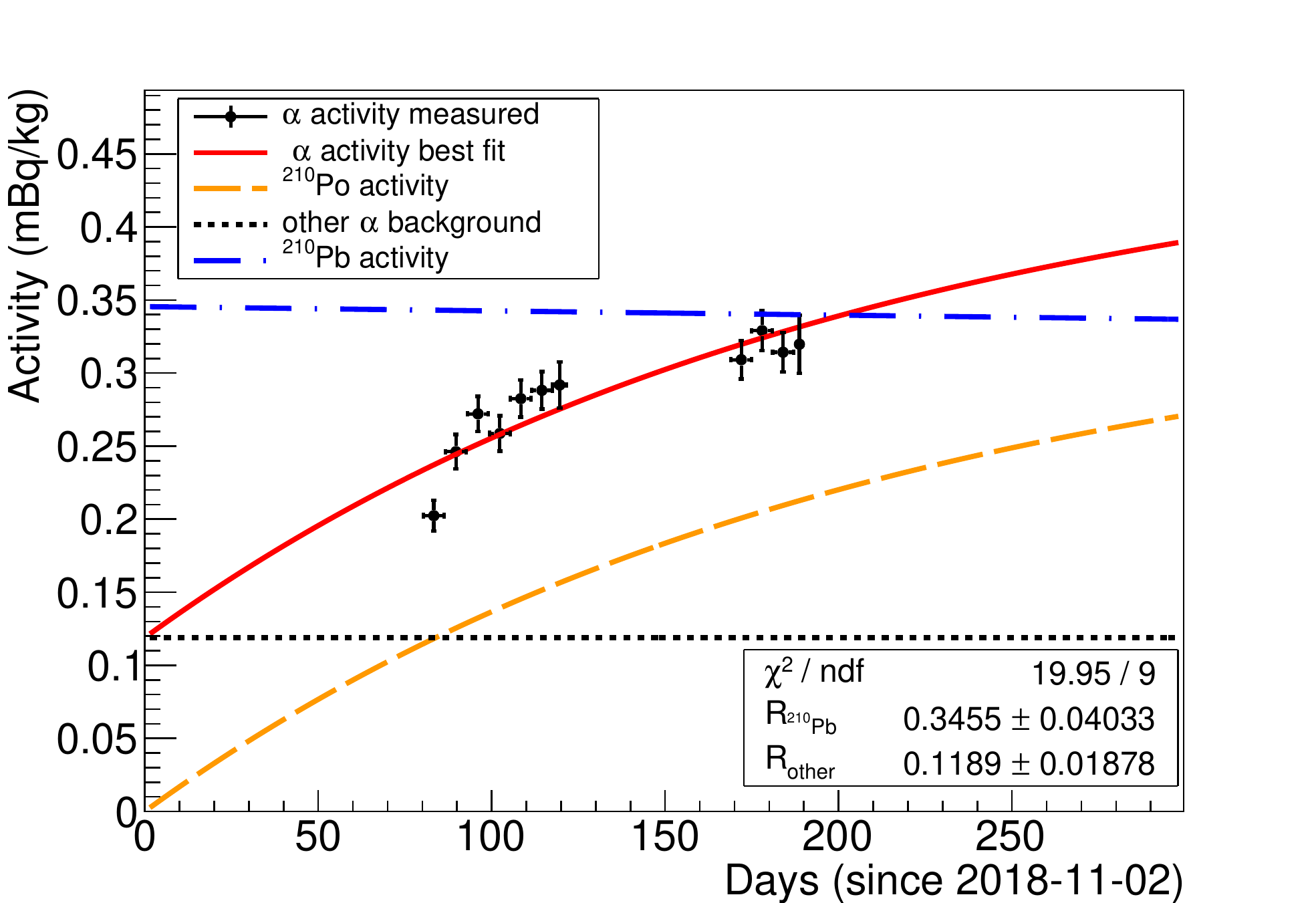}
    \else
        \includegraphics[width=0.7\linewidth]{figures/nai033_alpha_grp6_21.pdf}
    \fi
    \caption{\isotope{Po}{210} activity as a function of mean time of each group of data. The data points correspond to measured total $\alpha$ rates in the region of interest, and the red solid line the best fit to the total $\alpha$ rates, assuming a time-independent component due to background and a time-dependent component due to \isotope{Po}{210} ingrowth. The dotted black line and the dashed orange line illustrate the estimated constant background and the ingrowth component obtained from the best fit, respectively. The blue dot-dash line shows the expected activity of \isotope{Pb}{210} deduced from the activity of \isotope{Po}{210}. The activity of \isotope{Pb}{210} is estimated to be 0.34$\pm$0.04~mBq/Kg.}
    \label{fig:po-activity}
\end{figure}

The temporal behavior of $\alpha$ activity is compared to the decay chain, assuming a time-dependent \isotope{Po}{210} activity and a time-independent constant background~(Fig~\ref{fig:po-activity}). The best-fit indicates a \isotope{Pb}{210} activity of 0.34$\pm0.04$~mBq/kg. Although such a \isotope{Pb}{210} activity is high compared to 5-30~$\mu$Bq/kg in DAMA/LIBRA crystals~\cite{dl-apparatus}, it is lower than other current NaI-based dark matter experiments~\cite{anais,dmice, kims,cosine}.

\subsection{\isotope{H}{3}}

Another important intrinsic background source is \isotope{H}{3}, which can only be measured directly by scintillation counting. \isotope{H}{3} can be introduced into the final crystal as trace amount of water in the starting material~\cite{nai-thermal-peak}, or by cosmic activation. The former component is minimized by precision-drying of the powder prior to crystal growth~\cite{suerfu-thesis}. The production rate of cosmogenic \isotope{H}{3} from NaI is estimated to be 83$\pm$27~/d/kg~\cite{nai-cosmic} at sea level, corresponding to about 0.04~cpd/kg/keV background for each month of sea level exposure. NaI-033 has spent about 10~months on the surface before being transported to underground, \isotope{H}{3} background is thus expected to be $\approx$0.4~cpd/kg/keV.

\section{Discussions}

\isotope{K}{40} and \isotope{Pb}{210} are major sources of background for many NaI-based dark matter detectors. In NaI-033, the use of ultra-low potassium NaI powder and contamination-free growth technology makes \isotope{K}{40} no longer a dominant source of background.

Given the strict high-purity protocols we have adopted, it is likely that \isotope{Pb}{210} contamination comes from the raw material. However, the relative contributions to \isotope{Pb}{210} between the NaI powder and the TlI powder is not clear and requires an independent measurement.

Currently, NaI-033 is underground at LNGS for direct characterization of background. The background of NaI-033 in the SABRE setup is expected to be 0.8~cpd/kg/keV, calculated by scaling the results of~\cite{sabre-mc} except for \isotope{H}{3} where the result of \cite{nai-cosmic} and 10-month exposure to cosmic rays at sea level is assumed. For NaI crystals with dimensions similar to NaI-033, crystal growth alone takes 2 months. If surface exposure is minimized accordingly, the \isotope{H}{3} background could be reduced to 0.08~cpd/kg/keV and the overall background to 0.5~cpd/kg/keV. To further reduce \isotope{H}{3} background, crystal growth will have to be carried out underground.

\section{Conclusions}

In the quest for dark matter direct detection, the nature of the DAMA/LIBRA annual modulation has been a long-standing mystery. To make a model-independent test, NaI crystals of comparable or higher radiopurity than those used by DAMA/LIBRA are essential. In this article, we have presented our most recent ultra-high purity NaI crystal---NaI-033, featuring the lowest level of K ever achieved in NaI crystals and a relatively low level of \isotope{Pb}{210} among currently-running NaI(Tl) detectors. The growth of NaI-033 also demonstrates the effectiveness of the technologies and protocols we have developed in growing and processing ultra-high purity single crystals. These technologies and protocols are important not only for NaI crystals, but also for other single crystals for low-background applications.

\section*{Acknowledgements}

The research described in this article is the result of several years' painstaking research, trials and errors, and collaboration with industrial partners. The authors acknowledge Emily Shields, Jingke Xu and Francis Froborg for their foundational contributions without which this project would be impossible. The authors would like to thank Brad McKelvey at Seastar Chemicals for his indispensable role in developing key ICP-MS calibration and measurement techniques and crucible precision cleaning procedures. The authors would also like to thank Richard Morgan and Taylor Morgan at Sandfire Scientific for helping with crucible preparation. The authors would also like to acknowledge the help of and discussions with SABRE collaborators, especially A. Ianni, C. Vignoli, G. DiCarlo, S. Copello and A. Mariani. Burkhant Suerfu would like to thank Vinod Gupta for providing data storage and computational resources and Peter Meyers for carefully reviewing this article. Princeton group is supported by the National Science Foundation under award number PHY-1242625, PHY-1506397 and PHY-1620085. RMD group is supported by the Department of Energy under contract ID DE-SC0013760.


\bibliographystyle{apsrev4-1}
\bibliography{reference}

\end{document}